\documentclass[usenatbib,longtabl]{mn2e}

\title[W UMa contact binaries]
{The evolutionary status of W Ursae Majoris-type systems}

\author[Li et al.]{Lifang Li$^{1}$\thanks{E-mail:
llf@ynao.ac.cn or gssephd@public.km.yn.cn}, Fenghui Zhang$^{1}$,
Zhanwen Han$^{1}$, Dengkai Jiang$^{1,2}$, and Tianyu Jiang$^{1,2}$\\
$^{1}$National Astronomical Observatories, Yunnan Observatory,
Chinese Academy of Sciences, P.O. Box 110,
Kunming,\\
\ \ \ \  \ \ \ \ \ \ \ \ \ \ \ \ \ \ \ \ \ \ \ \ \ \ \ \ \ \ \ \ \
\ \ \ \  Yunnan Province, 650011, P.R. China\\
$^{2}$Graduate University of Chinese Academy Sciences, Beijing,
100039}
\begin{document}
\input psfig.sty
\date{Accepted 2007 November 15. Received 2007 November 7; in original form 2007 May 11}

\pagerange{\pageref{firstpage}--\pageref{lastpage}} \pubyear{2008}

\maketitle

\label{firstpage}

\begin{abstract}
Well-determined physical parameters of 130 W UMa systems have been
collected from the literature. Based on these data, the
evolutionary status and dynamical evolution of W UMa systems are
investigated. It is found that there is no evolutionary difference
between W- and A-type systems in $M-J$ diagram which is consistent
with the results derived from the analysis of observed spectral
type, $M-R$ and $M-L$ diagrams of W UMa systems. $M-R$ and $M-L$
diagrams of W- and A-type systems indicate that a large amount of
energy should be transferred from the more massive to the less
massive component so that they are not in thermal equilibrium and
undergo thermal relaxation oscillation (TRO). Meanwhile, the
distribution of angular momentum, together with the distribution
of mass ratio, suggests that the mass ratio of the observed W UMa
systems is decreased with the decrease of their total mass. This
could be the result of the dynamical evolution of W UMa systems
which suffer angular momentum loss (AML) and mass loss due to
magnetic stellar wind (MSW). Consequently, the tidal instability
forces these systems towards the lower q values and finally to
fast rotating single stars.

\end{abstract}

\begin{keywords}
binaries: close -- stars: magnetic fields -- stars: mass-loss--
stars: evolution
\end{keywords}

\section{Introduction}

W UMa stars are short-period, dumbbell-shaped binaries in which
both stars are in contact or overflowing their Roche limiting
surface. They are classified into A- and W-type systems on the
base of the light curve \citep{bin70}. A-type systems are showing
primary minima due to the eclipse of the larger, more massive
component while the reversal is true for W-type systems. However,
a satisfactory theory for the origin, structure and evolution of
the W UMa binaries has not yet been suggested. The traditional
view for the origin of contact binaries is that W UMa systems are
formed from detached binaries of comparable periods through
orbital decay by AML \citep{vil81,rah82, pac06}. \citet{pac06}
have found too many contact binaries in comparison with the
possible detached progenitors having orbital periods about one
day. Then, W UMa systems seem to appear out of "nowhere". Even if
it is possible that W UMa systems originate from detached
progenitors through AML due to MSW if they are formed from
detached binaries with periods $P\la$ 2.24 d (corresponding to a
maximum lifetime of about $\tau\la$ 3.23 Gyr in pre-contact stage)
and the lifetime of them are very long \citep{li07}. Otherwise,
one should not find so many contact binaries through ASAS. The
excess of W UMa stars implies other channel of formation.
Recently, \citet{pri06} have found that up to 59 per cent of W UMa
binaries have companions. This opens up a possibility that the
\citet{koz62} cycle operates in some triples. But, the relative
importance of the traditional origin of W UMa binaries through
AML, and through the Kozai cycle, will be required further study
\citep{pac06}. With regard to the lifetime of W UMa binaries, the
different lifetimes have been derived for W UMa systems by various
authors based on different materials, such as $0.1-1.0$ Gyr
\citep{gui88}, 1.61 Gyr \citep{bil05}, 4.7 Gyr \citep{van96}, 7.2
Gyr \citep{li05} and $\ga5.68$ Gyr \citep{li07}.

With regard to the evolutionary status of W UMa systems,
\citet{hil88} have argued that the primary components of the
shallow-contact W-type systems are unevolved main-sequence stars
while those of A-type ones are near to the terminal-age main
sequence. Meanwhile, there are other physical differences between
these two subclasses. A-type systems usually have an earlier
spectral type, a larger mass, a higher luminosity and a smaller
mass ratio. The degree of overcontact is larger and a thick common
envelope is present. W-type systems are showing a later spectral
type, a smaller mass, a lower luminosity, and a larger mass ratio.
They have a shallow convective envelope. So A- and W-type systems
are usually assumed to be in the different evolution schemes or
evolutionary status. However, the observed spectral type does not
indicate a difference in evolutionary state between A- and W-type
subgroups \citep[and references therein]{van82}. In addition, some
interesting ideas have been proposed concerning the evolutionary
link between A- and W-type systems. Some theories suggest the
possibility that W-type systems evolve into A-type through mass
exchange \citep{hil88,mac85} while others suggest the opposite
\citep{ruc85,gaz06}. Here three questions arise: do both subtypes
have the same genetic origin? regarding A types and W types, is
one type the progenitor of the other one and what will be the
final outcome of the evolution of each group? The above questions
are still open for investigations \citep{van82,gaz06}.

The components in a W UMa system have nearly equal surface
temperatures in spite of their often greatly different masses. The
canonical explanation of this phenomenon is that a large amount of
energy is transferred from the more massive to the less massive
component. However, the problem whether or not W UMa systems are
in thermal equilibrium also remained an open question. It is
interesting that there is controversy about it at present. Some
authors claim that a contact configuration can achieve thermal
equilibrium, such as \citet{ste06} argued that W UMa systems can
achieve thermal equilibrium if the secondary, which was the former
primary, is more advanced in its evolution, in analogy with the
`Algol paradox', so that W UMa systems should not undergo TRO.
However, it is not clear that this assertion is generally correct
\citep{pac07}. \citet{pac07} proposed that this is likely
applicable to systems with extreme mass ratios. In addition, other
authors claim that a contact configuration in thermal equilibrium
is not possible so that W UMa systems are forced to undergo TRO
\citep{li04,yak05,pac06}.

In this work, we have collected the well determined parameters of
130 W UMa systems (69 W- and 61 A-types) from the literature.
Based on these data, the evolutionary status and the dynamical
evolution of these binary stars are discussed in the following
Sections.

\section{Status of W UMa systems}

We collect the well-determined physical parameters of 78 W UMa
contact binaries from the compiles of \citet{yak05,awa05} and
\citet{mac96}. In addition, we collect the new physical parameters
for 52 W UMa systems from other literature (listed in Table 1).
Based on these data, the evolutionary status of W UMa systems has
been analyzed. The relations of $M-R$ and $M-L$ for W- and A-type
systems, together with the relations of $M-R$ and $M-L$ of zero
age main sequence (ZAMS) stars \citep{tou96}, are plotted in
Figures 1 and 2, respectively. It is seen in Figure 1a that most
of the primaries of W-type systems are above ZAMS and seem to have
evolved away from ZAMS, only the primaries of 13 W-type UMa
systems (GZ And, 44i Boo, AC Boo, EF Boo, VW Cep, TW Cet, TX Cnc,
SW Lac, ET Leo, ER Ori, BB Peg, HT Vir, and BD$+42^{\rm o}\ 2782$)
are under ZAMS. However, as seen from Figure 1b, apart from the
primaries of 13 W UMa systems mentioned above, the primaries of 27
other W-type systems are found to have moved under ZAMS.
Meanwhile, it is seen in Figure 2a that most of the primaries of
A-type systems are above ZAMS and seem to have evolved away from
ZAMS, only the primaries of 5 A-type systems (V417 Aql, BI CVn,
V899 Her, VZ Lib, and EQ Tau) are under ZAMS. But as seen from
Figure 2b, apart from the primaries of 5 W UMa systems mentioned
above, the primaries of 23 other A-type systems are found to have
moved under ZAMS. This indicates that the evolutionary status of
the primaries of the W UMa systems (including A- and W-types)
predicted by $M-R$ relation seem to be more advanced than that
predicted by $M-L$ relation. Meanwhile, both $M-R$ and $M-L$
diagrams show that the secondaries of A- and W-type systems appear
to have been evolved away from ZAMS. The appearance might be as a
result of the energy transfer from the primary to the secondary in
W UMa systems. According to \citet{haz77}, the effects of energy
transfer on the radii and effective temperatures of the primaries
of W UMa systems can be given by
\begin{equation}
\Delta{\rm log}R(t)=\int_0^t \psi_{_{\rm R}} (t-\tau)\dot E d\tau,
\end{equation}

\begin{equation}
\Delta{\rm log}T_{\rm e}(t)=\int_0^t \psi_{_{\rm T}} (t-\tau)\dot
E d\tau,
\end{equation}
where $\dot E$ is the rate of energy transfer, $\psi_{_{\rm R}}$
and $\psi_{_{\rm T}}$ are the response functions for the radius
and effective temperature with respect to energy transfer. The
luminosity $L \propto R^{2}T_{\rm e}^{4}$, the effect of energy
transfer on the luminosity can be given by
\begin{eqnarray}
\Delta{\rm log}L(t)&=&2\Delta{\rm log}R(t)+4\Delta{\rm log}T_{\rm
e}(t)\cr
 &=&\int_0^t (2\psi_{\rm R}+4\psi_{_{\rm T}}) (t-\tau)\dot
E d\tau.
\end{eqnarray}
The primaries of W UMa systems always lose the energy in the
common envelope, the energy loss would lead the effective
temperature and radius of the primaries to be deceased. According
Eqs. (1) and (2), $\psi_{_{\rm R}}$ and $\psi_{_{\rm T}}$ should
be positive \citep[also see][]{haz77}. Therefore, the response
function ($\psi_{_{\rm L}}=2\psi_{_{\rm R}}+4\psi_{_{\rm T}}$) for
the luminosity with respect to energy transfer is larger than that
($\psi_{_{\rm R}}$) for the radius, which implies that the
luminosity of the primary is affected by energy transfer more
greatly than its radius in the W UMa systems. Although the stellar
evolution would lead to continuous increase of luminosity and
radius, the energy transfer (loss) would lead their luminosities
and radii to be decreased and attempt to draw them back to ZAMS
line. Since the luminosity of the primary is affected by energy
transfer more greatly than its radius, the positions of the
primaries should be drawn back by energy loss more greatly in
$M-L$ diagram than in $M-R$ diagram. It is a main reason why $L$
values of the primaries appear to be lower than ZAMS line while
$R$ values of the primaries are higher than those of the systems
on ZAMS. This suggests that the status of the primaries of W UMa
systems indicated by $M-R$ diagram seems to be more advanced than
that predicted by $M-L$ diagram.

Figures 1 and 2 also show that both radius and luminosity of the
secondary of each W UMa system are evidently deviated from those
of ZAMS stars, and that the evolutionary status of the secondaries
appears to be more advanced than that of the primaries. Most
probably, this must be the result of the energy transfer from the
primary to the secondary in the W UMa systems. In general, the
energy transferred from the primary to the secondary in the
envelope of W UMa systems is severe or even tens of times more
than the nuclear evolutionary effect on the secondaries so that
the positions of the secondaries are significantly different from
those of ZAMS stars.

\begin{figure}
\centerline{\psfig{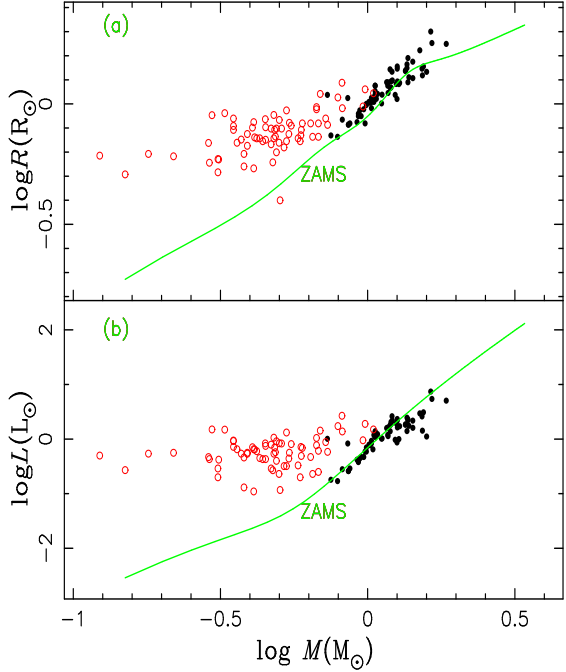}}
\caption{Relations of mass-radius and mass-luminosity for W-type W
UMa systems. Open and full dots represent the less massive and
more massive components, respectively. The solid lines represent
ZAMS from \citet{tou96}.}
\label{fig1}
\end{figure}

\begin{figure}
\centerline{\psfig{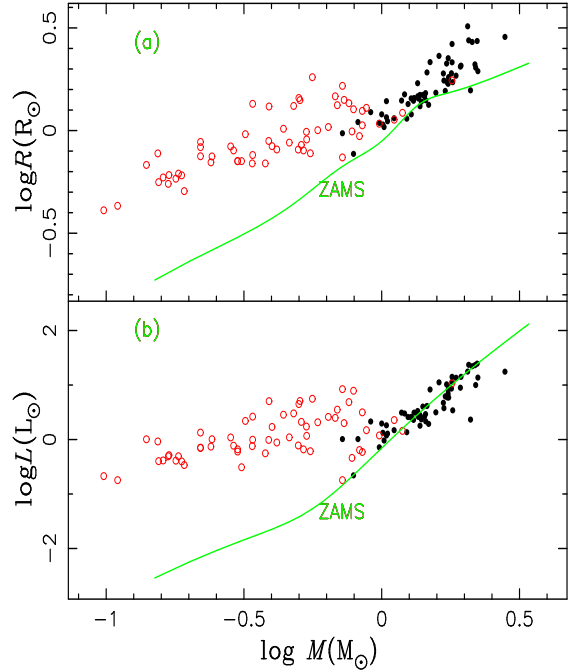}}
\caption{Relations of mass-radius and mass-luminosity for A-type W
UMa systems. Solid lines and symbols are the same as Fig. 1.}
\label{fig2}
\end{figure}

\section{The possible outcome of W UMa systems}

\begin{table*}
\begin{footnotesize}
Table~1.\hspace{4pt} New physical parameters of some contact binaries.\\
\begin{minipage}{17cm}
\begin{tabular}{l|cccccccccc}
\hline\hline\
{Stars}&{Type}&{$P$}&{$M_{1}$}&{$M_{2}$}&$R_{1}$& $R_{2}$&{$L_{1}$}&{$L_{2}$}&{$q_{_{\rm ph}}$}&{References}\\
&&{(days)}&{($M_{\rm \odot}$)}& {($M_{\rm \odot}$)}&{($R_{\rm
\odot}$)}
&{($R_{\rm \odot}$)}&{($L_{\rm \odot}$)}& {($L_{\rm \odot}$)}&&\\
\hline
AB And &W&0.3319&1.042&0.595&1.025&0.780&0.648&0.492&0.571&1\\
GZ And &W&0.3050&1.115&0.593&1.005&0.741&1.017&0.717&0.532&1\\
V417 Aql&A&0.3703&1.377&0.498&1.314&0.808&1.796&0.777&0.355&1\\
V402 Aur&W& 0.6035&1.638 & 0.327 & 1.997& 0.951&7.425&1.491&0.199&1 \\
DN Cam&W&0.4983&1.849&0.818&1.775&1.224&5.062&2.668&0.442&1\\
YY CrB&A&0.3766&1.393&0.339&1.385&0.692&2.347&0.755&0.232&1\\
SX Crv &A& 0.3166 &1.246 & 0.098 &  1.347 & 0.409&2.590&0.213&0.0787&1 \\
V2150 Cyg&A & 0.5919 &2.233 &1.798 & 1.946 & 1.756&13.707&10.721&0.802&1 \\
GM Dra &W& 0.3387 &1.213 &0.219 & 1.252 &0.606&2.190&0.562&0.210&1 \\
UX Eri&A&0.4453&1.430&0.534&1.468&0.905&2.637&1.169&0.13&1\\
V829 Her&W&0.3582&0.856&0.372&1.058&0.711&0.829&0.541&0.435&1\\
SW Lac&W&0.3207&1.240&0.964&1.090&0.976&0.971&0.953&0.787&1\\
AP Leo &A&0.4304&1.359&0.416&1.433&0.809&2.596&0.882&0.297&1\\
AO Cam&W&0.3299&1.119&0.486&1.092&0.732&1.029&0.574&0.435&1\\
VZ Lib &A& 0.3583 &1.480 & 0.378& 1.335 & 0.692&1.934&0.559&0.255 &1 \\
DZ Psc &A &0.3661 & 1.352& 0.183& 1.469& 0.617&2.836&0.493&0.145&1\\
QW Gem&W&0.3581&1.262&0.413&1.239&0.726&1.633&0.645&0.334&1\\
GR Vir &A& 0.3470 &1.376 & 0.168 & 1.490 &0.550&2.806&0.493&0.106&1\\
NN Vir &A&0.4807&1.730&0.850&1.717&1.246&5.905&3.155&0.61&1\\
DK Cyg&A&0.4707&1.741&0.533&1.708&0.986&8.157&1.731&0.306&1\\
AH Aur&A&0.4943&1.674&0.283&1.897&0.837&4.729&1.090&0.165&1\\
EF Boo&W&0.4295&1.547&0.792&1.431&1.064&3.084&1.731&0.534&1\\
HN UMa &A&0.3826&1.279&0.179&1.435&0.583&2.55&0.41&0.147&2 \\
EQ Tau&A&0.3413&1.233&9.551&1.143&0.775&1.36&0.61&0.447&2\\
FU Dra&W&0.3067&1.173&0.312&1.110&0.588&1.13&0.42&0.256&2\\
BB Peg&W&0.3615&1.424&0.550&1.279&0.813&1.61&0.81&0.386&2\\
OU Ser&A&0.2968&1.109&0.192&1.148&0.507&1.48&0.34&0.172&2\\
V776 Cas&A&0.4404&1.750&0.242&1.821&0.748&5.90&1.01&0.138&2\\
UV Lyn&W&0.4150&1.344&0.501&1.376&0.858&1.86&0.84&0.372&2\\
V592 Per&A&0.7157&1.743&0.678&2.252&1.468&9.58&2.50&0.389&2\\
HT Vir&W&0.4077&1.284&1.046&1.223&1.107&1.72&1.50&0.815&2\\
CK Boo&A&0.3352&1.442&0.155&1.521&0.561&2.924&0.401&0.106&3\\
FP Boo&A&0.6405&1.604&0.154&2.310&0.774&11.193&0.920&0.096&3\\
XZ Leo&A&0.4877&1.742&0.586&1.689&1.004&6.926&2.073&0.336&3\\
ET Leo&W&0.3465&1.586&0.542&1.359&0.835&1.115&0.564&0.342&3\\
AQ Psc &A& 0.4756 &1.682&0.389&1.753&0.890&3.760&0.984&0.231&3\\
V921 Her&A&0.8774&2.068&0.505&2.752&1.407&23.526&5.094&0.244&3\\
V839 Oph&A&0.4090&1.572&0.462&1.528&0.874&3.148&1.097&0.53&3\\
V2357 Oph&W&0.4156&1.191&0.288&1.392&0.689&1.782&0.468&0.231&3\\
VY Sex&W&0.4434&1.423&0.449&1.497&0.864&2.174&0.832&0.315&3\\
FI Boo&A&0.3900&0.82&0.31&1.10&0.71&1.02&0.31&0.382&4\\
XY Boo&A&0.3706&0.912&0.169&1.230&0.607&2.138&0.515&0.1855&5\\
V523 Cas&W&0.2337&0.75&0.38&0.74&0.55&0.18&0.13&0.512&6\\
V781 Tau&W&0.3449&1.29&0.57&1.212&0.852&1.39&0.70&0.405&7\\
TV Mus&A&0.4457&1.35&0.22&1.70&0.83&3.33&0.71&0.166&8\\
BD+42 2782&W&0.3702&1.38&0.67&1.29&0.95&1.75&1.05&0.482&9\\
DX Tuc& A&0.3771&1.00&0.30&1.20&0.71&1.97&0.66&0.29&10\\
V870 Ara&W&0.3997&1.503&0.123&1.67&0.61&2.96&0.50&0.082&10\\
SS Ara&W&0.4060&1.343&0.406&1.369&0.797&1.83&0.71&0.302&11\\
AH Cnc&W&0.3605&1.21&0.18&1.36&0.62&2.62&0.54&0.610&12\\
V899 Her&A&0.4212&2.100&1.190&1.570&1.220&2.320&1.44&0.566&13\\
U Peg&W&0.3748&1.17&0.39&1.23&0.75&1.19&0.54&0.324&14\\
 \hline
\end{tabular}
\end{minipage}
\end{footnotesize}\\
{References in Table 1: (1) Gazeas et al. 2005; (2) Zola et al.
2005; (3) Gazeas et al. 2006; (4) Terrell et al. 2006;(5) Yang et
al. 2005; (6) Zhang \& Zhang 2004; (7) Yakut et al. 2005; (8) Qian
et al. 2005;(9) Lu et al. 2007, (10) Szalai et al. 2007; (11) Lu
1991; (12) Zhang, Zhang \& Deng 2005; (13) \"Ozdemir et al. 2002;
(14) Djura\v svi\'c et al. 2001}
\end{table*}

The angular momentum distribution of 78 W UMa systems had been
discussed by \citet{mac96}. It is found that most of A-type
systems seem to have no evolutionary link with W-type ones because
A- and W-type systems can be divided into two separate parts in
$M-J$ diagram by a line which gives the angular momentum of a
system with just in contact main sequence components and a primary
mass of 1.35 $M_{\rm \odot}$. The orbital angular momentum of a
binary can be written as
\begin{equation}
J_{\rm orb} = \frac{M_{1}M_{2}}{M}A^{2}\omega_{\rm o},
\end{equation}
where $M$ and $M_{1,2}$ are the total mass and the masses of both
components, respectively, $A$ the orbital separation and
$\omega_{\rm o}$ the orbital angular velocity. The spin angular
momentum of the binary is
\begin{equation}
J_{\rm s}=(k_{1}^{2}M_{1}R_{1}^{2}
+k_{2}^{2}M_{2}R_{2}^{2})\omega_{\rm s}.
\end{equation}
where $R_{1,2}$ are the radii of the primary and secondary,
$k_{1,2}$ the ratios of the gyration radius to the stellar radius
for both stars and $\omega_{\rm s}$ the spin angular velocity. In
general, the rotational periods of the components are synchronized
with the orbital periods for W UMa systems, due to strong tidal
interaction (i.e. $\omega_{\rm s}=\omega_{\rm o}=\omega$). If the
ratios of the gyration radius to the stellar radius are assumed to
be equal (i.e. $k_{1}^{2}=k_{2}^2=k^{2}$), the total angular
momentum of the binary is
\begin{equation}
J=\bigl\{\frac{M_{1}M_2}{M}A^{2}+k^{2}(M_1R_{1}^{2}+M_2R_{2}^{2})\bigr\}\omega.
\end{equation}

Taking $k^{2}=0.06$ as in \citet{ras95} and \citet{li06}, the
angular momentum distribution of 130 W UMa systems with
well-determined parameters is investigated. A very interesting
feature is shown in Figure 3, where the total angular momentum of
130 observed W UMa systems based on equation (6) is plotted $vs$
the total mass with full and open dots for A-types and W-types,
respectively. Figure 3 confirms the fact that the total angular
momentum ($J$) decreases by decreasing total total mass ($M$) of W
UMa binaries.

In order to obtain the information on the dynamical evolution of W
UMa systems from the angular momentum distribution, it is
necessary to give a theoretical distribution of angular momentum
$vs$ the total mass for W UMa systems. If $\omega$ is in units of
day$^{-1}$, $M$ in $M_{\rm \odot}$ and $A$ in $R_{\rm \odot}$, the
Kepler's third law can be written as
\begin{equation}
\omega=54.23(\frac{M}{A^{3}})^{1/2},
\end{equation}
the total angular momentum of the binary reads
\begin{equation}
J=54.23M^{3/2}A^{1/2}\mu\Bigl\{(1-\mu)+k^2r_{1}^{2}\bigl(1+\frac{1-\mu}{\mu}(\frac{r_{2}}{r_{1}})^{2}\bigr)\Bigr\},
\end{equation}
in which
\begin{equation}
\mu=M_{1}/M,
\end{equation}
where $\mu$ is a mass fraction of the primary of the system, and
$r_{1,2}$ the relative radii of both components. According to
\citet{egg83,egg06}, they can be approximately written as
\begin{equation}
r_{1}\simeq\frac{R_{\rm L1}}{A}
=\frac{0.49\bigl\{(1-\mu)/\mu\bigr\}^{-2/3}}{0.6\bigl\{(1-\mu)/\mu\bigr\}^{-2/3}+{\rm
ln}\Bigl\{1+\bigl\{(1-\mu)/\mu\bigr\}^{-1/3}\Bigr\}},
\end{equation}
\begin{equation}
\frac{r_{2}}{r_{1}} \simeq\bigl(\frac{1-\mu}{\mu}\bigr)^{0.46}.
\end{equation}
According to \citet{bin70}, the contact condition can be expressed
as
\begin{equation}
A=\frac{R_{1}+R_{2}}{0.76}=1.32R_{1}\bigl\{1+(\frac{1-\mu}{\mu})^{0.46}\bigr\}.
\end{equation}

Mass-radius relation for the primary of W- and A-type systems can
be expressed as
\begin{equation}
R_{1}=c_{\rm r}M_{1}^{\beta}=c_{\rm r}\mu^{\beta}M^{\beta}.
\end{equation}
Inserting the equations (10)--(13) into equation (8), one can
obtain
\begin{equation}
J=62.31c_{\rm r}^{1/2}f(\mu)M^{(3+\beta)/2},
\end{equation}
in which
\begin{eqnarray}
f(\mu)&=&\mu^{(2+\beta)/2}\bigl\{1+(\frac{1-\mu}{\mu})^{0.46}\bigr\}^{1/2}\Bigl\{(1-\mu)+\cr
&&k^{2}r_{1}^2\bigl\{1+(\frac{1-\mu}{\mu})^{1.92}\bigr\}\Bigr\},
\end{eqnarray}
then
\begin{equation}
{\rm log}J=\frac{3+\beta}{2}{\rm log}M+c(\mu).
\end{equation}
Based on Eqs. (14) and (16) in \citet{awa05}, $\beta$ can be
derived to be 0.62 for A- and W-type systems, and $c_{\rm r}$ can
be derived to be 1.05 (for W-types) or 1.23 (for A-types). Then,
$c_{\rm r}$ is taken to be 1.23 since A-types are located near the
two boundaries of $M-J$ diagram, and $\mu$ is taken to be 0.5
($q=1.0$) and 0.93 ($q=0.075$), respectively. Equation (16) is
also shown in Figure 3 as two dashed lines with a slope of 1.81.
It is seen in Figure 3 that all observed W UMa systems are located
in a strip limited by the two dashed lines and the observed
systems with extreme mass ratios, such as V870 Ara, CK Boo, FP
Boo, V776 Cas, SX Crv, FG Hya, DZ Psc, AW UMa, GR Vir are on or
near the lowest boundary \citep[corresponding to a mass ratio
close to a cutoff mass ratio of 0.076 for W UMa systems
in][]{li06}. The mass ratio of W UMa systems is decreased from the
upper boundary to the lower boundary.

\begin{figure}
\centerline{\psfig{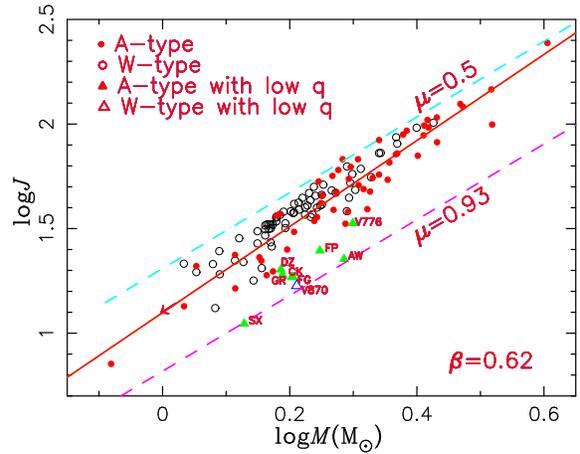}}
\caption{Total angular momentum of W UMa systems based on equation
(6) as a function of total mass. Dashed lines represent the
theoretical angular momentum distributions [based on equation
(16)] of W UMa systems with mass ratio of 1.0 and 0.075,
respectively. Solid line represents a fitting result of
observations, i.e. equation (17).}
\label{fig3}
\end{figure}

\citet{awa05} have computed only the orbital angular momentum
(OAM) of contact binaries and plotted it on a $M-J$ diagram, then
fit a quadratic function to it. However, a linear function can fit
the distribution of the total angular momentum $vs$ mass in their
logarithmic scale for the observed W UMa systems well in present
work, and a least-squares solution results a relationship between
the total angular momentum and mass as the following

\begin{equation}
{\rm log}J=2.06(8){\rm log}M+1.10(2),
\end{equation}
with correlation coefficient $r=0.91$ and standard deviation
$sd=0.10$. Equation (17) is also plotted in Figure 3 as a solid
line with a higher slope than those of the two dashed lines based
on Equation (16). This suggests that the mass ratio might be
decreased with the decrease of the total mass for the observed W
UMa systems.

\begin{figure}
\centerline{\psfig{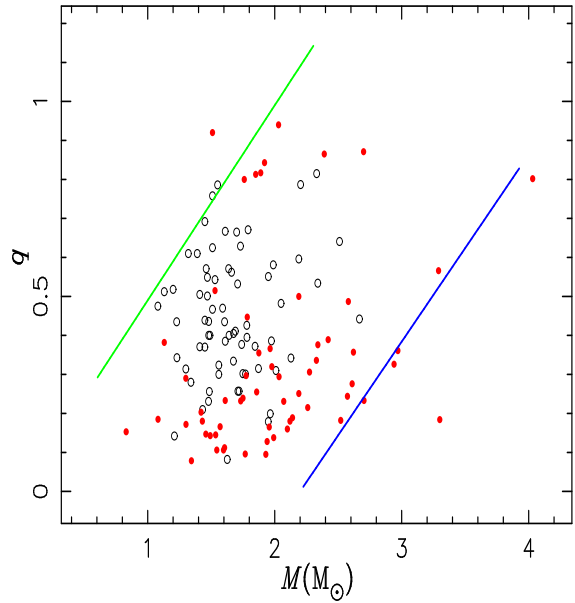}}
\caption{Mass ratio of W UMa systems as a function of total mass.
Full and open dots represent the A-type and W-type systems,
respectively.}
\label{fig4}
\end{figure}

The distribution of mass ratio $vs$ the total mass for the above
systems is shown in Figure 4. It is seen in Figure 4 that the W
UMa systems seem to be located in a strip and there is a tendency
for a decreasing total mass $M$ with decreasing mass ratio $q$,
which is similar to the result given by \citet{van96}. Meanwhile,
as seen from Figure 4, the mass ratios of W-type systems are
located in a region from 0.3 to 0.7 and the mass ratios of A-type
systems seem to be located in two separated regions ($q\la0.5$ and
$q\ga$ 0.7). Meanwhile, it is seen in Figure 3 that there is no
obvious difference between the angular momentum distributions of
W-types and A-types. This is in good agreement with a result that
the observed spectral type does not indicate a difference in
evolutionary state between A- and W-type systems \citep{van82},
and A- and W-type systems are probably in the different stages of
TRO.

The difference in the slope of equations (16) and (17) can be
attributed to many causes, such as rough assumptions, mean values
of coefficients taken from empirical mass-radius relations,
observational selection effects, initial mass and period
distributions, and all effects including the effect of dynamical
evolution. The rough assumptions and the coefficient $c_{_{\rm
r}}$ might only change the location (up or down) of the
theoretical distribution given by equation (16) in Figure 3,
rather than its slope. The slope of equation (16) is only
influenced by the coefficient $\beta$ of empirical mass-radius
relation. However, equation (16) with a higher slope than equation
(17) requires $\beta>$ 1.0 which is much larger than a value
derived by \citet{awa05}. \citet{ruc06} has found that the contact
binaries discovered by ASAS lacks those with photometric
amplitudes $<0.4$ mag. This might imply that a telescope has a
detection threshold and some W UMa systems with smaller mass
ratios will be difficult to be detected because of the very small
amplitude of the corresponding light curves \citep{van78,ruc01}.
Furthermore, spectroscopic mass ratios were always determined for
the best observable double lined spectroscopic W UMa systems. This
selection favours the high mass ratio binaries \citep{van78}.
Figures 3 and 4 show that the W UMa systems with smaller total
masses usually have smaller mass ratios. Therefore, the
observational selection effects possibly lead the slope of
equation (17) to be decreased. It is believed that the W UMa
systems are formed from detached binaries through AML due to MSW
or through \citet{koz62} cycle \citep{egg01,pac06}. Then, the
initial mass and period distribution would affect the time
dependent formation process of these systems (i.e. the lifetime in
the pre-contact stage). The lifetime in the pre-contact would
affect mass, mass ratio, and nuclear evolutionary degree of W UMa
contact binaries just formed from the detached ones if their
progenitors suffer mass loss and AML owing to MSW. The W UMa
systems which originate from detached or semi-detached ones with
main sequence components and short periods, usually spend a short
time in pre-contact and have a high mass ratio of about 0.7
\citep{moc81,li04,li05}. W UMa contact binaries with a relatively
long lifetime in pre-contact are usually formed from the detached
ones which have a long period and a small radius ratio of the
components to their Roche lobes. Since the more massive component
evolves faster than the less massive one, the radius ratio of the
secondary to its Roche lobe remains a relatively low value when
the primary begins to fill its Roche lobe so that the more mass is
required to transfer from the primary to the secondary in the
duration from the semidetached binary to a W UMa contact binary,
with result that W UMa contact binaries should have a higher mass
ratios. In addition, a lower relative energy transfer rate is
required for W UMa systems with a higher evolutionary degree
\citep{liu00}, the lower energy transfer rate should cause a
smaller expansion of the radius of the secondary. In this case
only the systems with a higher mass ratio might satisfy the
requirement of Roche geometry. Therefore, W UMa systems with a
relatively long lifetime in pre-contact should have a higher mass
ratio and less mass. This would also lead the slope of equation
(17) to be decreased.

A higher slope of equation (17) compared with equation (16) is
probably caused by the dynamical evolution of the W UMa systems.
In fact, it is consistent with not only the prediction of the
theoretical models \citep{luc76,fla76,rob77,li04,li05} but also
the observational mass ratio distributions given by \citet{van78}
and \citet{ruc01}, who have argued that the W UMa systems with low
mass ratios are much more than those with high mass ratios. If
this is true, the W UMa systems would merge into the fast-rotating
stars within $10^{3}-10^{4}$ yrs when their mass ratios is
decreased to the cutoff mass ratio (0.076 \citealt{li06}, 0.09
\citealt{ras95}). Furthermore, there is another possibility as
suggested by \citet{pac07} that W UMa type binaries, especially
for those with extreme mass ratios have secondaries in an advanced
evolutionary stage as Algol-type ones. However, the secondary with
a much larger radius than it would be on the main sequence is not
only a result of its advanced nuclear evolution but also a result
of energy transfer from the primary to it. If so, W UMa systems
with extreme mass ratios and those with high mass ratios should
have the different formation processes.

\section{Discussions and conclusions}

The status of the primaries of W UMa systems (including A- and
W-type systems) predicted by $M-R$ diagram seems to be more
advanced than that indicated by $M-L$ diagram. Meanwhile, the
primaries are almost located on ZAMS line for A- and W-type
systems in Figures 1b and 2b. The appearance might be caused by
the energy transfer from the primary to the secondary in W UMa
systems, and the luminosity of the primaries of W UMa systems is
affected by energy loss more greatly than their radii. In
addition, the secondaries are significantly deviated from ZAMS
stars in $M-R$ and $M-L$ diagrams. This might also be the result
of a large amount of energy transfer from the more massive to the
less massive component in W UMa systems.

Since a large amount of energy is transferred from the primary to
the secondary in W UMa systems, both components of W UMa systems
can not achieve thermal equilibrium. At the beginning of contact
stage, the efficiency of energy transfer is extremely low since
the common envelope is too thin to undertake so much energy
transfer, with the result that the mass is still transferred from
the primary to the secondary as in semidetached stage. The
decrease of orbital separation due to mass transfer should lead to
the continued increase of the contact depth and energy transfer.
The added mass on to the secondary would stop when the significant
energy is transferred from the primary to the secondary. This
followed by mass transfer back to the primary until the decrease
in radius of the secondary caused by mass loss can not compensated
by its increase caused by energy transfer. Once this has occurred,
the full efficiency of energy transfer is lost. Then, the
secondary contracts and the contact depth decreases rapidly
although the mass is transferred from the primary to the secondary
in this stage. The secondary breaks contact and continues to
collapse towards a main sequence equilibrium configuration, with
its temperature and luminosity falling rapidly. The luminosity of
the primary decreases during the semidetached phase, since the
Roche lobe again contracts to prevent its free expansion towards
thermal equilibrium. Since the process of raising matter up
through the star, for transfer to the secondary requires
significant quantities of energy, at the expense of the surface
luminosity. Meanwhile, the timescales of the primary and the
secondary are very different, because of mass accretion of the
secondary its radius will increase before it collapses to thermal
equilibrium configuration. Therefore, the two components of W UMa
systems are unlikely to be in thermal equilibrium, but each star
attempts to reach thermal equilibrium. The attempt of the W UMa
systems to reach a non-existent thermal equilibrium, coupled with
Roche geometry, would force W UMa systems to undergo TRO on a
period of about thermal timescale of the primary \citep[and
references therein]{li04}. Although the primaries of W UMa systems
do not reach thermal equilibrium, they are near to the thermal
equilibrium and exhibit the property that the luminosity is
affected by energy transfer more greatly than its radius. But, the
thermal timescale of the secondaries is much longer than the
cyclic period so that the secondaries are evidently deviated from
thermal equilibrium.

Although Figures 1a and 2a show that the stage of evolution of
A-type systems seems to be higher than that of W-type ones as
suggested by \citet{luc79}. However, these objects are still
within the main sequence band and we can hardly can speak of
"evolved" objects, and this kind of binaries should last a very
long time \citep{van82}. Based on a simple assumption that the
primaries of A- and W-type systems are ZAMS stars, according to
\citet{csi04}, the nuclear luminosity ratio of the secondary to
the primary can be approximately expressed as
\begin{equation}
\frac{L_{2}}{L_{1}}=(\frac{M_{2}}{M_{1}})^{4.6}=q^{4.6}.
\end{equation}
Since the two components of W UMa systems have almost equal
effective temperatures, i.e. $T_{1}\simeq T_{2}$, the observed
luminosity ratio of them reads
\begin{equation}
\frac{L_{2,{\rm observed}}}{L_{1,{\rm
observed}}}=\frac{L_{2}+\Delta L}{L_{1}-\Delta L
}=(\frac{R_{2}}{R_{1}})^{2}=q^{0.92},
\end{equation}
where $\Delta L$ is the energy transfer rate between the
components. Insert equation (19) into equation (20), one can
obtain a relative energy transfer rate as
\begin{equation}
\frac{\Delta L}{L_{1}}=\frac{q^{0.92}-q^{4.6}}{1+q^{0.92}},
\end{equation}
the relation of the relative energy transfer rate {\it vs} mass
ratio is shown in Figure 5. It is seen in Figure 5 that the
relative energy transfer rate of W UMa systems peaks at a mass
ratio of about 0.55 which is just located in the region of mass
ratio distribution of W-types, whilst the mass ratios of most
A-types is significantly deviate from the maximum energy transfer
rate of W UMa systems. Therefore, if the energy transfer is taken
in account, we hardly can draw such a conclusion that A-types are
evolved more greatly than W-types so that A- and W-type systems
have the the different evolutionary status. This confirms the
suggestions originally made by \citet{van82}, which there is no a
difference in evolutionary state between A- and W-type systems.
Combining with the different mass ratio distributions of A- and
W-type systems, we conclude that A- and W-type systems might be in
the different stages of TRO, i.e. A-type systems might be in the
stages form contact evolution to semi-detached one or from
semi-detached configuration to contact one. This is exactly
consistent with a relatively high temperature difference between
the two components in A-type systems. However, no a TRO model can
predict the observed number ratio of A-type systems with lower
mass ratios to those with higher mass ratios. This suggests that
present treatments of the energy transfer might be inadequate.

\begin{figure}
\centerline{\psfig{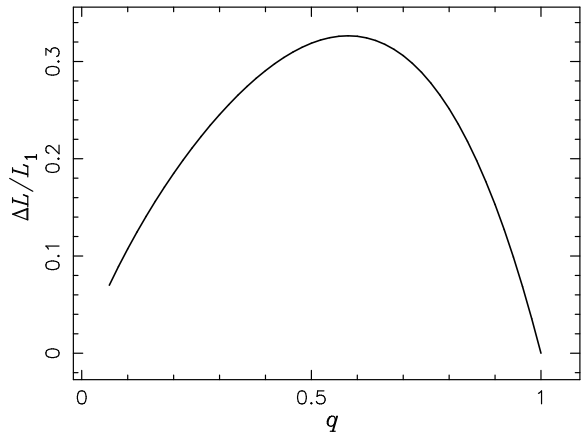}}
\caption{Relative energy transfer rate between both components as
a function of mass ratio of W UMa systems.}
\label{fig6}
\end{figure}

The distribution of angular momentum and mass ratio suggests that
the smaller is the total mass of the observed W UMa systems, the
smaller is their mass ratios. It might be caused by the dynamical
evolution of the W UMa systems, or by mass ratio reversal of the
progenitors of W UMa systems \citep{ste06}, especially for those
with extreme mass ratios \citep{pac07}. In addition, the first
possibility is consistent with the present theoretical models
\citep{luc76,fla76,rob77,li04,li05} and the observed mass ratio
distribution \citep{van78,ruc01}, i.e. the dynamical evolution
would lead W UMa systems to evolve into those with smaller mass
ratios, then lead them to merge into fast-rotating stars owing to
tidal instability. However, the second possibility might exist and
it indicates that W UMa systems with extreme mass ratios and those
with high mass ratios might have different formation processes.
Meanwhile, the relative importance of these two possibilities
requires further study through more observations.

At present, it is not clear whether or not there is an
evolutionary link between A- and W- type systems. If there is an
evolutionary link between A- and W-type systems, based on the
evolutionary tendency of mass ratio of W UMa systems, we can
conclude that A-type systems with relatively high mass ratios
would probably evolve into W-type systems, then evolve into A-type
ones with relatively low mass ratios, and the opposite direction
seems to be impossible, because it requires some W UMa systems
with mass ratios close to 1 and different temperature between the
two components. However, up to now there are in total of four
contact binaries which are found to have mass ratios close to 1,
they are V701 Sco \citep{bel87}, CT Tau \citep{ple93}, V803 Aql
\citep{sam93}, and WZ And \citep{zha06}. Moreover, only a system
WZ And in which the components with almost equal mass have
different temperature between the components is found, it might be
unequal mass system through mass transfer and mass reverse. In
each of the other three systems, the components with equal mass
have almost identical temperature between the components. They
could be regarded as congenital twins, rather than an outcome due
to mass transfer during the evolution.

\section*{ACKNOWLEDGEMENTS}

The authors gratefully acknowledge an anonymous referee for many
insightful comments and ideas about contact binaries, which
improved the paper greatly. This work was partly supported by the
Chinese Natural Science Foundation (10673029, 10773026, and
10433030), the Foundation of Chinese Academy of Sciences
(KJX2-SW-T06), and by the Yunnan Natural Science Foundation
(2007A113M and 2005A0035Q).

\bsp

\label{lastpage}

\end{document}